\documentstyle[12pt,aps,prc,epsfig,preprint]{revtex}
\tightenlines
\begin{document}
\begin{titlepage}

\title{Unified optical-model approach to low-energy antiproton annihilation
on nuclei and to antiprotonic atoms}
\author{C.J. Batty$^a$, E. Friedman$^b$, A. Gal$^b$ \\
$^a$Rutherford Appleton Laboratory, Chilton, Didcot, Oxon OX11 0QX, UK \\
$^b$Racah Institute of Physics, The Hebrew University, Jerusalem
91904, Israel}
\maketitle

\begin{abstract}
	A successful unified description of $\bar p$ nuclear interactions
        near $E=0$ is achieved using a $\bar p$ optical potential within
        a folding model, $V_{{\rm opt}} \sim \bar v * {\rho}$,
        where a $\bar p p$ potential $\bar v$ is folded with the nuclear
        density $\rho$. The potential $\bar v$ fits very well the
        measured $\bar p p$ annihilation
        cross sections at low energies ($p_L < 200$ MeV/c) and the $1s$ and
        $2p$ spin-averaged level shifts and widths 
        for the $\bar p$H atom. The density-folded optical
        potential $V_{{\rm opt}}$ reproduces satisfactorily the
        strong-interaction level shifts and widths over the entire periodic
        table, for $A > 10$, as well as the few 
        low energy $\bar p$
        annihilation cross sections measured on Ne. Both $\bar v$ and
        $V_{{\rm opt}}$ are found to be highly absorptive, which leads to
        a saturation of reaction cross sections in hydrogen and on nuclei.
        Predictions are made for $\bar p$ annihilation cross sections
        over the entire periodic table at these very low energies and the
        systematics of the calculated cross sections as function of $A$, $Z$
        and $E$ is discussed and explained in terms of a Coulomb-modified
        strong-absorption model. Finally, optical potentials which fit
        simultaneously low-energy $\bar p - ^4$He observables for $E < 0$
        as well as for $E > 0$ are used to assess the reliability of
        extracting Coulomb modified $\bar p$ nuclear scattering lengths
        directly from the data. The relationship between
        different kinds of scattering lengths is discussed and
        previously published  systematics
        of the $\bar p$ nuclear scattering lengths is updated.
\newline
$PACS$: 24.10.Ht, 25.43.+t, 25.60.Dz
\newline
 \newline
%\vspace{1cm}
Corresponding author: E. Friedman\newline
Tel: +972 2 658 4667,
Fax: +972 2 658 6347, \newline
E mail: elifried@vms.huji.ac.il
\end{abstract}
\centerline{(\today )}
\end{titlepage}

\section{Introduction}
\label{sec:int}

The present paper is motivated by the recent publications of experimental
annihilation cross sections for antiprotons on several targets at very
low energies ($p_{L} < 100$ MeV/c) \cite{ZBB99a,ZBB99b,BBB00a}.
Some unexpected features observed in the dependence of
these cross sections on the mass ($A$) and charge ($Z$) numbers have been very
recently explained by us \cite{GFB00} as arising from saturation
of the $\bar p$ annihilation cross sections due to
the strong absorption that is characteristic of the $\bar p$-nucleus
interaction. Using an optical model approach we have shown that the $E=0$
borderline between $\bar p$ atoms and the scattering regime can be crossed
using the same $\bar p$-nucleus potential,
i.e. one can get good fits to experimental results both
for $\bar p$ atoms and for low energy annihilation cross sections with the
same potential, at least within a given mass region of the periodic table.
Such potentials are strongly absorptive, which leads to
a remarkable saturation of the total reaction cross section with
increasing $A$. Strong absorption has very recently been shown
\cite{FGa99a,FGa99b} to lead also to saturation
of the widths of $\bar p$ atomic states and to the prediction of relatively
narrow deeply bound $\bar p$ atomic states. The close
analogy between bound-state widths and total reaction cross sections
is best demonstrated by observing the corresponding expressions,
assuming for simplicity a Schr\"odinger-type equation.
The width $\Gamma$ of an atomic level is then given in terms of the
optical potential $V_{\rm opt}$ by:

\begin{equation} \label{equ:gamma}
\frac{\Gamma}{2}= -\frac{\int {\rm Im} V_{{\rm opt}}(r)
| \psi({\bf r}) | ^2  d {\bf r}}
{\int | \psi({\bf r}) | ^2  d {\bf r}}\quad.
\end{equation}
Here $\psi({\bf r})$ is the $\bar p$ full atomic wavefunction.
The corresponding expression for the total reaction cross section
at positive energies is

\begin{equation} \label{equ:sigma}
\sigma_R = -\frac{2}{\hbar v} \int
{\rm Im} V_{{\rm opt}}(r) | \psi({\bf r}) |^2 d {\bf r}\quad.
\end{equation}
Here $\psi({\bf r})$ is the $\bar p$-nucleus elastic scattering
wavefunction and $v$ is the c.m. velocity. We note that these
expressions involve no approximation.

In the present work we extend the earlier work by presenting a full
analysis of the dependence of calculated
annihilation cross sections on $A$, on $Z$ and on the energy
up to 16 MeV ($p_{L} = 175$ MeV). Beginning with the $\bar p p$ system,
we show that the available experimental results, including the
strong-interaction spin-averaged shift and widths of the $\bar p$H atom,
are very well accounted for by a potential
approach. Moving over to $\bar p$-nuclear systems, we observe that the
strong-interaction level shifts and widths in $\bar p$ atoms are well
reproduced for $A > 10$, over the entire periodic table, by folding the
above $\bar p p$ potential $\bar v$ with the nuclear density and
renormalizing slightly its strength, to obtain a $\bar p$-nucleus optical
potential. Using this density-folded optical potential $V_{\rm opt}$,
we study the systematics of the predicted cross sections over the entire
periodic table and show it to result naturally from the absorptive
properties of the interaction under conditions of Coulomb focussing.
The optical potentials are then used to derive
$\bar p$-nucleus $s$-wave scattering lengths. We study the $A$ dependence
of these scattering lengths and compare them with those extracted
directly from the low-energy data using approximation methods \cite{PBL00}.

The paper is organized as follows. In Section \ref{sec:opt} we define
the optical potential and the various scattering amplitudes which are used
throughout the present work.
Section \ref{sec:H} deals with the $\bar p p$ system and in Section
\ref{sec:heavy} we discuss the $\bar p$ interaction with nuclei.
Section \ref{sec:He} is devoted to the  $\bar p$ - $^4$He system
which is used for closely examining the scattering length approximation
and Section \ref{sec:SLheavy} extends the discussion to heavier
systems. The work is summarized and concluded in Section \ref{sec:summ}.

\section{Optical potentials and scattering amplitudes}
\label{sec:opt}

The interaction of low energy antiprotons
with nuclei, as well as the interaction of antiprotons bound
in an atomic system, is described in this work by the conventional
`$t\rho$' potential \cite{BFG97}
\begin{equation}\label{eq:potl}
2\mu V_{{\rm opt}}(r) =
 -{4\pi}(1+{\frac{A-1}{A}\frac{\mu}{m}})b_0\rho(r) \;\;\; ,
\end{equation}
where $m$ is the nucleon mass, $\mu$ is the $\bar p$-nucleus reduced
mass, $b_0$ is an `effective scattering length' complex parameter
obtained from fits to the data and $\rho(r)$ is the nuclear density
distribution normalized to $A$. The density $\rho(r)$ may also include
the effect of folding in a finite-range two-body interaction.
The factor $(A-1)/A$ in Eq. (\ref{eq:potl}) is usually omitted in
large $A$ atomic studies \cite{BFG97}, as we do also here
(Sect. \ref{sec:heavy}), but is retained of course in the discussion
of the $\bar p p$ system in Sect. \ref{sec:H} and the $\bar p$-He system
in Sect. \ref{sec:He}.

It is customary to describe the interaction of bound hadrons
with the nucleus by the Klein-Gordon (KG)
 equation of the form:

\begin{equation}\label{eq:KG1}
\left[ \nabla^2  - 2{\mu}(B+V_{{\rm opt}} + V_c) + (V_c+B)^2\right] \psi = 0~~ ~
~
(\hbar = c = 1) \;\;\; ,
\end{equation}
where $B$ is the complex binding energy
 and $V_c$ is the finite-size
Coulomb interaction of the hadron with the nucleus, including
vacuum-polarization terms.
Equation (\ref{eq:KG1}) assumes that $V_{{\rm opt}}$ behaves
as a Lorentz scalar.
The use of the KG equation for $\bar p$ atoms is  justified as long as
spin effects are negligible and one is interested in ($2j+1$) averaging.
As we are interested in this work in crossing the $E$=0 borderline between
the bound atomic system and the low energy scattering regime, we use the same
wave equation also for positive energies, namely

\begin{equation} \label{eq:KG2}
\left[ \nabla^2 + k^2 - (2\varepsilon^{(A)}_{red}V_c -
V^2_c) -2 \mu V_{{\rm opt}} \right] \psi = 0~~ ~~(\hbar = c = 1)
\end{equation}

\noindent
where $k$ and $\varepsilon^{(A)}_{red}$ are the
wave number and reduced energy respectively in the c.m.
system, and $(\varepsilon^{(A)}_{red})^{-1}=E_{\bar p}^{-1}+E_A^{-1}$ in
terms of the c.m. total energies for the projectile and target,
 respectively.
Equation (\ref{eq:KG2}) is the Lorentz scalar version of the KG equation
used in Ref.\cite{FGM97} for studying $K^+$ interaction with nuclei.
Near $E=0$ both equations yield practically the same numerical results. Furthermore,
we have verified that using the Schr\"odinger equation instead
of the KG equation
leads to calculated reaction cross sections that differ by 0.5\%
or less.

The scattering amplitude due to the wave equation (\ref{eq:KG2}) can be
written in the form \cite{Schiff}
\begin{equation}
\label{eq:ampl1}
f(\theta) = f^{(c)}(\theta) + e^{2i\sigma}f^{(sc)}(\theta) \quad ,
\end{equation}
where $f^{(c)}(\theta)$ is the point Coulomb scattering amplitude,
\begin{equation}
\label{eq:amplc}
f^{(c)}(\theta) = \frac{1}{k}\sum (2l+1) C^{(c)}_l
P_l({\cos}\theta) \quad , \quad\quad
C^{(c)}_l =  e^{i\sigma _l} {\sin}\sigma _l \quad ,
\end{equation}
and with non-relativistic Coulomb phase shifts given by
\begin{equation}
\label{eq:sigC}
\sigma _l = {\arg} \Gamma(1+l-i\eta) \quad ,
\quad\quad \eta = \frac{1}{ka_B} \quad ,
\end{equation}
with $a_B=\hbar ^2 /(Z\mu e^2)$ the Bohr radius.
If $f(\theta)$ is defined to have the phase shifts
$\sigma _l + \delta ^{(sc)}_l$, with Coulomb-modified
phase shifts $\delta ^{(sc)}_l$ due to the short-ranged
part of the interaction, then the partial-wave expansion 
 of the Coulomb-modified scattering
amplitude $f^{(sc)}(\theta)$ is given by
\begin{equation}
\label{eq:amplS}
f^{(sc)}(\theta) =  \frac{1}{k}\sum (2l+1) C^{(sc)}_l
P_l({\cos} \theta) \quad , \quad\quad
C^{(sc)}_l =  e^{i\delta ^{(sc)}_l} {\sin} \delta ^{(sc)} _l \quad ,
\end{equation}
subject to the following partial-wave representation of the operator
product $e^{2i\sigma}f^{(sc)}(\theta)$:
\begin{equation}
\label{eq:operS}
e^{2i\sigma}f^{(sc)}(\theta) = \frac{1}{k}\sum (2l+1)
e^{2i\sigma _l} C^{(sc)}_l P_l({\cos} \theta) \quad .
\end{equation}
In practice, the Coulomb-modified partial wave amplitudes $C^{(sc)}_l$
are determined by solving the wave equation for each partial wave with
point Coulomb boundary conditions at sufficiently large distance.
The total reaction cross section is then given by
\begin{equation}
\label{eq:sigR}
\sigma_R = \frac{4 \pi}{k^2} \sum (2l+1) [{\rm Im} C^{(sc)}_l -
({\rm Im} C^{(sc)}_l)^2
- ({\rm Re} C^{(sc)}_l)^2] \quad .
\end{equation}

In addition to the $\bar p$ reaction cross section, which near $E = 0$
is  essentially exhausted by the total annihilation cross section,
we also discuss in this paper the low energy effective-range expansion
which requires the knowledge of the Coulomb-modified
phase shifts $\delta ^{(sc)}_l$.
These are given in terms of the partial wave amplitudes as follows:
\begin{equation}
\label{eq:Imdelta}
{\rm Im} \delta ^{(sc)}_l = - \frac{1}{4} {\ln}
[(1 - 2 {\rm Im} C^{(sc)}_l)^2 + 4 ({\rm Re} C^{(sc)}_l)^2] \quad ,
\end{equation}
\begin{equation}
\label{eq:Redelta}
{\cot} ({\rm Re} \delta ^{(sc)}_l) = i \frac{[1-2 {\rm Im} C^{(sc)}_l +
2 i {\rm Re} C^{(sc)}_l]
e^{2 {\rm Im} \delta ^{(sc)}_l} +1}
{[1-2 {\rm Im} C^{(sc)}_l + 2 i {\rm Re} C^{(sc)}_l]
e^{2 {\rm Im} \delta ^{(sc)}_l} - 1} \quad ,
\end{equation}
where the latter expression is particularly useful in order to avoid a
possible ambiguity in $\delta ^{(sc)}_l$ if it is determined from
an alternative expression involving ${\sin}({2 \rm Re}~ \delta ^{(sc)}_l)$.
We note that for absorptive potentials $V_{{\rm opt}}$, the
Coulomb-modified phase shifts are complex, with Im~$\delta ^{(sc)}_l \ge 0.$
An alternative expression for $\sigma _R$ in terms of transmission
coefficients $T_l$ is then given by
\begin{equation}
\label{eq:sigmaR2}
\sigma _R = \frac{\pi}{k^2} \sum (2l+1) T_l \quad, \quad\quad
T_l = 1 - e^{-4 {\rm Im} \delta ^{(sc)}_l} \quad ,
\end{equation}
which explicitly shows that $\sigma _R$=0 for real potentials.

Finally, the low-energy effective-range expansion for $l = 0$, in
the presence of an attractive Coulomb potential, is given by
\cite{Bethe49,Tru61}:

\begin{equation}
C_0^2\left( \eta  \right)k\cot \delta _0^{\left( {sc} \right)}-{2
\over {a_B}}h\left( \eta  \right)=-{1 \over {a_0^{\left( {sc} \right)}}}+{1
\over 2}r_0^{\left( {sc} \right)}k^2+O\left( {k^4} \right) \  ,\  \
	\label{eq:range1}
\end{equation}
where
\begin{equation}
C_0^2\left( \eta  \right)={{2\pi \eta } \over {1-e^{-2\pi
\eta }}}\  , \qquad
h\left( \eta  \right)= {\rm Re}~  \psi \left( {i\eta } \right)-\ln \eta \
,\  \ 	\label{eq:gamow1}
\end{equation}
and $\psi$ is the digamma function \cite{ASt65}. For $\eta {_{_{_{_>}}}
\atop ^{^\sim}} 1$, or equivalently $ka_{B} {_{_{_{_<}}} \atop ^{^\sim}}
1$, the following limits hold:

\begin{equation}
C_0^2( {\eta {_{_{_{_>}}} \atop ^{^\sim}}1})\approx 2\pi
\eta \  ,\ \qquad h\left( \eta  \right) \approx {1 \over {12\eta ^2}}+{1 \over
{120\eta ^4}}+\ldots \  ,\  \
      \label{eq:gamow2}
\end{equation}
so that Eq. (\ref{eq:range1}) for the effective range expansion reduces
to
\begin{equation}
{{2\pi } \over {a_B}}\cot \delta _0^{(sc)}\approx -{1 \over {a_0^{\left( {sc}
\right)}}}+{1 \over 2}\left( {{1 \over 3}a_B+r_0^{\left( {sc} \right)}}
\right)k^2+O\left( {k^4} \right) \  .\  \	\label{eq:range2}
\end{equation}
In the present work we solve numerically the scattering wave equation
down to $E_{L} = 100$~keV, where the linear dependence of the l.h.s.
on $k^{2}$ is explicitly verified, in order to extract reliably the
Coulomb modified low energy parameters. We recall, that in the absence
of a Coulomb potential, the effective range expansion for a
short-ranged potential (such as $V_{\rm opt}$) assumes the form:

\begin{equation}
k\cot \delta _0\approx -{1 \over {a_0}}+{1 \over 2}r_0k^2+O\left( {k^4}
\right)\   ,	\label{eq:range3}
\end{equation}
which can be derived from Eq. (\ref{eq:range1}) upon taking the
limit $\eta \to 0$. The Coulomb-modified scattering length
$a_{0}^{(sc)}$ is related to the purely short-ranged scattering length
$a_{0}$ by the following approximate expression \cite{JBl50,Hol99}:

 \begin{equation}
{1 \over {a_0^{\left( {sc} \right)}}}\approx {1 \over {a_0}}+{2 \over
{a_B}}\left( {\ln {{a_B} \over {2R}}+1-2\gamma } \right)\   , 	\label{eq:JBl}
 \end{equation}
 which assumes that $R \ll a_{B}$, where $R$ is the range of the
 short-ranged potential and $\gamma = 0.5772\dots$ is the Euler
 constant.

\section{The antiproton-hydrogen system}
\label{sec:H}

As a first step towards studying the $\bar p$-nucleus system
we tried to fit the data for the $\bar p$
hydrogen  system with as
simple an `optical potential' $\bar v$ as possible, choosing a
Gaussian shape common to both real
and imaginary parts. In the present application
to the $\bar p p$ system the `density' $\rho(r)$
in Eq. (\ref{eq:potl}) was chosen as proportional to a Gaussian
${\exp}(-r^{2}/a_{G}^{2})$
and  normalized to a volume integral of 1. Total
reaction cross sections were calculated at 6 momenta
between 38 and 175 MeV/c \cite{ZBB99a,Ber96}
and the complex parameter $b_0$ was varied in order to fit
these recent low-energy data for a given value of $a_G$.
This was repeated for several values of $a_G$ between 1 and 2 fm.
In all cases excellent fits to the annihilation cross sections were achieved,
and consequently the range parameter $a_G$ could not be determined uniquely
from the annihilation cross sections. As we aim at using the same
potential on both sides of the $E=0$ borderline, we turned next to
the $\bar p$H atom.
 Using the
same type of   potential as above,  excellent
agreement between calculation and experiment \cite{AAB99,GAA99} was achieved
by varying $b_0$ for any
value of $a_G$ between 1 and 2 fm.
 However, requiring
that  both atomic and scattering data be described by the same
potential $\bar v$, we find $a_G$=1.5 fm and
$\bar b_0=-0.15+i1.8 $ fm, leading to
a combined $\chi ^2$ of 3.8 for the 6 data points for annihilation
cross sections
plus the 3 data points for the atom.
The uncertainties are $\pm0.15, \pm0.15$ and $\pm0.06$ fm for $a_G$,
Re $\bar b_0$ and Im $\bar b_0$, respectively.
This potential will be referred to
as potential (G). A more elaborate potential was previously
shown  \cite{Bru86,BBB00b} to be able to fit the
low-energy $\bar p p$ annihilation data.
 In that potential the real
part is described by a Saxon-Woods (SW) shape with a radius of 1.89 fm
and diffuseness parameter of 0.2 fm whereas the imaginary part has the SW
shape with a radius of only 0.41 fm and the same diffuseness of 0.2 fm.
Adopting these geometries we repeated the procedure described above
for fitting the annihilation and atomic data and obtained very good fits
with a $\chi ^2$ of 5.8 for the 9 data points. The parameter $\bar b_0$
is $2.85+i16.5$ fm, with about $\pm$10\% uncertainties,
leading to potential depths of 46.5 and 7550 MeV for the real
and imaginary parts, respectively, in very good agreement
with Refs. \cite{Bru86,BBB00b}. This potential will be referred to as
potential (SW). We note that for both potentials (G) and (SW),
the imaginary part is
considerably stronger than the real part, which signals a
very strong absorptivity
in the $\bar p p$ system. The real part of the potential,
under such circumstances,
plays  only a minor role \cite{GFB00}.
The imaginary parts of the potentials (G) and (SW) have markedly
different ranges. The very short range of (SW) is qualitatively
in agreement with other widely used phenomenological
potentials \cite{DR80,KW86}, whereas the range of (G) is about
twice as large as for these potentials.

\begin{figure}
\epsfig{file=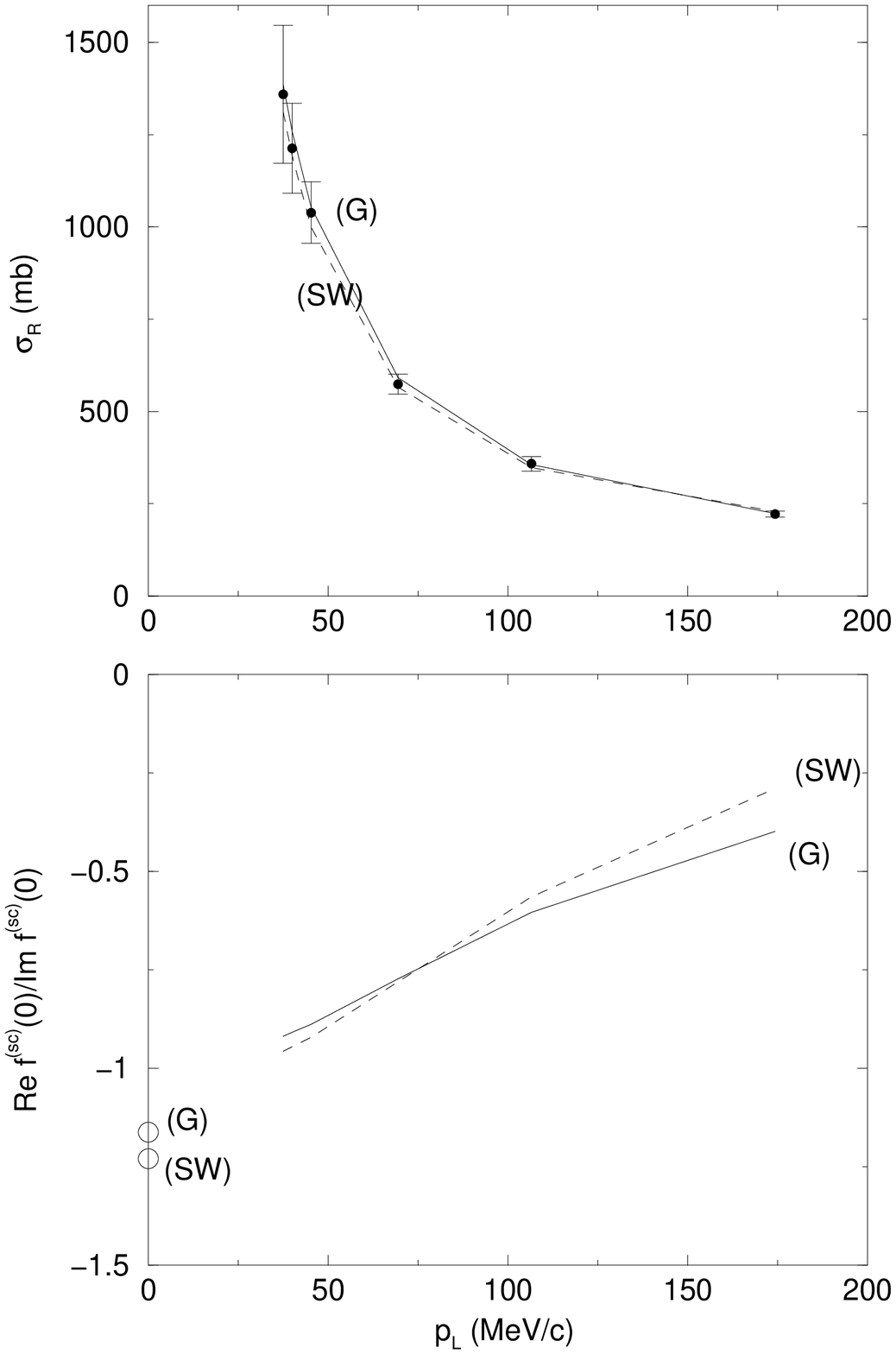,height=180mm,width=135mm,
bbllx=62,bblly=51,bburx=491,bbury=699}
\caption{Top: experimental annihilation cross sections 
[1,16] for the
$\bar p p$ system and calculated values using the (G) and (SW) potentials.
Bottom: ratios of the
real part of the $\bar p p$ Coulomb-modified forward scattering amplitude
to its imaginary part for the same potentials. Real to imaginary ratios for
the zero-energy scattering length are also shown.}
\label{fig:H}
\end{figure}

Figure \ref{fig:H} shows results for the $\bar p p$ system.
The upper part compares calculated reaction cross sections
for the two potentials with the experimental values, and the
excellent agreement is evident. The lower part shows the
ratios of real to imaginary parts of the forward
Coulomb-modified scattering amplitude $f^{(sc)}$
[see Eqs. (\ref{eq:ampl1}) - (\ref{eq:operS})] for the two
potentials, which is a quantity of considerable interest:
this, so called $\rho$ parameter, was determined at higher
$\bar p$ incident momenta to be positive, tending to zero
between  200 - 300 MeV/c \cite{BDG85}. Our calculations give
negative values for $\rho$ at this lower-momentum range,
indicating a smooth transition to the corresponding ratios for
the Coulomb-modified scattering length also shown in the lower
part, at zero energy, for potentials (G) and (SW).
Evidently the two potentials,
in spite of their different geometry, are very similar in predicting
observable quantities.

\section{The antiproton-nucleus system}
\label{sec:heavy}

With an established $\bar p p$ interaction potential $\bar v$,
the simplest way of incorporating this potential into heavier
$\bar p$-nuclear systems would be to fold it with the nuclear density
to obtain a $\bar p$-nucleus potential:
$V_{\rm opt} = \bar v * \rho$. Setting aside possible
renormalization effects in nuclei, this folding procedure also tacitly
assumes that at low energy the spin-averaged $\bar p n$ interaction
is approximately equal to the spin-averaged $\bar p p$ interaction,
as supported by the near equality of the imaginary parts of the
corresponding scattering lengths \cite{MCK88} and also by several
phenomenological $\bar N N$ potentials \cite{DR80,KW86}. However, recent
$\bar n p$ low-energy annihilation data from the OBELIX collaboration
\cite{BBB97} suggest that the $\bar n p$ interaction,
which by charge symmetry is equal to the $\bar p n$ interaction,
is considerably weaker than the $\bar p p$ interaction. Therefore,
we  keep the strength of the density-folded $\bar p$-nucleus
potential as a free parameter in our search, as is discussed below.

In the energy range covered by the present work the majority of
data come from $\bar p$ atoms and therefore this naive approach
is best tested by applying the resulting potentials to $\bar p$ atoms.
We emphasize that the $\bar p$ - nucleus potentials obtained
from $\bar p$ atoms are not unique. For `macroscopic' nuclear
densities of the type discussed in Ref. \cite{BFG97},
and for $A>10$,
a Gaussian with a range parameter of about 1.5 fm leads to
the best fit. 
Folding in the $\bar pp$ (G) potential with `macroscopic' nuclear
densities, 
we find that a {\it density-folded} optical potential of the form
(\ref{eq:potl}), with $b_0 = -0.1 +i1.2$ fm, fits all $\bar p$
atoms heavier than boron ($A>10$) with a $\chi ^2$ of 2.4 per point,
whilst the corresponding value for a best-fit phenomenological potential
of the form Eq. (\ref{eq:potl}) using the same unfolded 
nuclear densities is 2.7 \cite{BFG97}. We denote this $\bar p$-nucleus
density-folded optical potential by (F). The value of its strength
parameter $b_0$ is 2/3 of the strength $\bar b_0$ of the $\bar p p$
potential (G). 
Note that this is not a best-fit potential but merely a renormalized
G $*$ $ \rho$ potential, where a common factor is applied to the real
and imaginary parts.
Turning to $E>0$, accurate experimental annihilation
cross sections at low energies are scarce in this mass range,
and we only note that the calculated cross sections for Ne are in
reasonably good agreement with the data \cite{BBB00a,BBB86}, as can be
seen from Table \ref{tab:Ne}.

Having gained confidence in the applicability of the density-folded
optical potential (F) for nuclei, we use it
to calculate $\bar p$ reaction cross sections over the periodic table
for a range of energies, in order to study the systematics of the dependence on
$Z$, on $A$ and on the energy. Figure \ref{fig:heavy} shows, in the upper part,
the calculated reaction cross sections for $\bar p$ on Ne, Ca, Zr, Sn and Pb
at 37.6, 57 and 106.6 MeV/c. It is
seen that the range of values is very broad; for a given energy the
calculated cross sections vary by a factor of between 6 and 10. For a
given nucleus the cross sections vary
with energy by a factor between 3.2 and 5.7.
The mechanisms causing these variations are obviously of interest.

For strongly absorbed particles such as low energy antiprotons,
disregarding the Coulomb potential, the total
reaction cross section may be approximated by $\pi R^2$ where $R$ is
the radius of the nucleus.
The underlying assumption is that antiprotons with impact parameter
less than $R$, or equivalently with orbital angular momentum up to
$l_{\rm max}$ where semiclassically $l_{\rm max} + 1/2 = kR$, are
totally absorbed. However, due to the focussing effect of the
attractive Coulomb potential, particles with impact parameters
larger than $R$ also interact with the nucleus,
thus causing the cross section to increase. The low energy
Coulomb problem is well described by the semiclassical
approach \cite{Schiff} as recognized a long time ago by Blair in
connection with $\alpha$ particle reactions on nuclei \cite{Bla54}.
Assuming that total absorption occurs in all partial waves for
which the distance of closest approach is smaller than $R$, one
gets the following relation between the Coulomb-modified $l_{\rm max}$
and $R$:

\begin{equation} \label{eq:lmax}
(l_{\rm max} + \frac {1} {2})^2 \approx (k R)^2
(1 + \frac {2 \eta} {k R}) \;\;\; .
\end{equation}
At very low energies, $2\eta >> kR$, and therefore $l_{\rm max} >> kR$
due to the focussing effect of $V_c$. The total reaction cross section
in the strong absorption limit is then given by

\begin{equation}  \label{eq:sigma}
\sigma_R = \frac{\pi}{k^2} \sum (2l+1) \approx \frac{\pi}{k^2}
(l_{\rm max} +\frac{1}{2})^2 \approx \pi R^2 (1 + \frac{2 \eta}{k R})
= \pi R^2 (1 + \frac{2 m Z e^2}{\hbar^2 k_L k R}) \;\;\; ,
\end{equation}
where $k_L$ and $k$ are the laboratory and c.m. wave numbers, respectively.
The second term within the brackets represents the Coulomb
focusing effect and
at very low energies it becomes dominant \cite{GFB00},
thus leading to an $A^{1/3}Z$ dependence of the cross section
if  $R=r_0A^{1/3}$. At high energies the usual strong
absorption value $\pi R^2$ is obtained.
In order to use this expression one needs to define an equivalent radius $R$
that will properly represent the $\bar p$-nucleus interaction at the
particular energy. We proceed to define such a radius.

Expanding the $\bar p$-nucleus scattering wave function in partial waves,
\begin{equation}
\psi({\bf r})=\sum (2l+1)\frac{\psi_l(r)}{r} P_l({\cos}\theta) \;\;\; ,
\end{equation}
we define an average radius $<r_l>$ for the $l$ partial wave
\begin{equation}
<r_l>=\frac{\int r W(r) |\psi_l(r)|^2 dr}{\int W(r)|\psi_l(r)|^2 dr}
\end{equation}
with $W(r)=-$Im $V_{{\rm opt}}(r)$.
Then, bearing in mind the partial-wave expansion of the reaction
cross section Eq. (\ref{eq:sigmaR2}), we define an average
radius as follows:
\begin{equation}
<r> = \frac{\sum (2l+1) T_l <r_l>}{\sum (2l+1) T_l} \;\;\; .
\label{eq:ravg1}
\end{equation}
Expanding alternatively the reaction cross section Eq. (\ref{equ:sigma})
in partial waves:
\begin{equation}
\sigma _R = \frac{2}{\hbar v}
\int W(r) 4\pi \sum (2l+1) |\psi _l (r)|^2 d r \;\;\; ,
\end{equation}
one finds that
\begin{equation}
T_l = \frac{8 \mu k}{\hbar^2} \int W(r) |\psi_l(r)|^2 dr \;\;\; ,
\end{equation}
which combined with Eq. (\ref{eq:ravg1}) leads to an equivalent expression
for $<r>$:
\begin{equation} \label{eq:avgr}
<r> = \frac{\int W(r)| \psi({\bf r}) | ^2 r  d {\bf r}}
{\int W(r)| \psi({\bf r}) | ^2  d {\bf r}} \;\;\; .
\end{equation}
Note that the average radius turns out to be slightly energy dependent.
At $p_L$ = 37.6 MeV/c, the lowest $\bar p$ incident momentum for which
$\bar p p$ annihilation has been measured \cite{ZBB99a}, the calculated
values of $<r>$ appropriate to the density-folded optical potential (F)
can be parameterized to better than 1\% by
\begin{equation} \label{eq:rAdep}
<r> = 1.840 + 1.120A^{1/3} \; {\rm fm} \;\;\; .
\end{equation}
In fact, this expression holds to better than 2\% for the whole range
of very low energies up to $p_L \sim 100$ MeV/c.

In the lower part of Fig. \ref{fig:heavy} are plotted the ratios of the
calculated cross sections to the semiclassical expression (\ref {eq:sigma})
where for $R$ we took either the  radius of a rigid sphere having the same average
radius as the optical potential (F), namely $R=\frac{4}{3}<r>$, or just
$R=<r>$ which could be more appropriate due to the substantial attenuation of the
wavefunction which is involved in the definition of $<r>$.
It is seen that these ratios vary remarkably little, by less than 5\%
for the lower band, when the cross sections
vary by factors between 3 and 10. At the lower energy the Coulomb focusing
is dominant but at the higher energies
covered here both terms of Eq. (\ref{eq:sigma}) are important. It may therefore
be concluded that the dependence of the calculated reaction cross sections
on the three parameters ($Z, A$ and energy) is almost exclusively
given by Eq. (\ref{eq:sigma}).

Another consequence of the strongly absorptive
$\bar p$-nucleus optical potential is the saturation of
the physical observables which are induced by the
absorptivity. Thus, for $E < 0$, widths of $\bar p$
atomic levels saturate as function of the magnitude of
Im $V_{\rm opt}$ \cite{FGa99a,FGa99b} and, for $E > 0$,
total reaction cross sections also saturate with it
\cite{GFB00}. As argued in our earlier work \cite{GFB00},
the saturation property of $\sigma _R$ for any given
nucleus also implies that $\sigma _R$ does not rise with
$Z$ and $A$ as fast as it would have risen if the $\bar p$
absorption were considerably weaker and therefore could
be treated perturbatively. This is demonstrated here
differently, in Fig. \ref{fig:sat}, where the saturation
factor $S$ defined as
\begin{equation}
\label{eq:sat}
S = \frac{\sigma_R ({\rm Im}b_0) / {\rm Im}b_0}
{\sigma_R (10^{-4}{\rm Im}b_0) / 10^{-4}{\rm Im}b_0}
\end{equation}
is plotted at $p_L = 57$ MeV/c (1.73 MeV)
as function of $A$, using the
density-folded optical potential (F) for $A > 10$, (F')
for $^4$He (see below in Sect. \ref{sec:He})
and (G) for H. For Im $b_0$ as small as
$10^{-4}$ of its nominal value (of 1.2 fm for potential (F)),
the calculated $\sigma_R$ is very nearly linear in Im $b_0$.
If the linear rise of $\sigma_R$ were sustained up to
the nominal value, the saturation factor $S$ would have assumed
the value $S=1$. However, for Im $b_0$ = 1.2 fm for potential (F),
the calculation is strongly nonperturbative, as witnessed
in the figure by the small values of $S$ relative to 1,
about 0.1 and less beginning with Ne.
The saturation of $\sigma_R$ has two
aspects to it: (i) that $S$ depends very weakly on Im $b_0$
in the nonperturbative regime (not shown in the
figure); and (ii) that $S$ monotonically decreases with $A$,
as is shown in the figure, both for antiprotons (solid curve)
and for antineutrons (dashed curve). The values of $S$
for antiprotons are lower than for antineutrons due to the
Coulomb focussing effect which for antiprotons acts in two
seemingly opposing ways: (i) enhancing the reaction cross section
in the negatively charged case compared to the uncharged case,
but also (ii) expelling more effectively
\cite{FGa99a,FGa99b} the $\bar p$ wavefunction from the
nuclear region, so that $\sigma_R$ becomes
more strongly suppressed than for antineutrons. This latter
effect gets weaker with increasing energy; at $E_L = 6.04$ MeV,
$S_{\bar p}$ is larger by 20 - 25 \% than at 1.73 MeV, whereas
$S_{\bar n}$ is smaller than at 1.73 MeV.

\section{The antiproton - helium system}
\label{sec:He}

In this section we study the $\bar p$ - $^{3,4}$He system at very low
energies, within the optical model approach, in order to test the
scattering-length approximation (SLA) which was successfully applied
to the $\bar p p$ system \cite{CPZ97} and has recently been extended
to the $\bar p d$ and $\bar p$ - $^4$He systems \cite{PBL00}.
$^4$He is one of the very few nuclei for which
data on the strong interaction of
$\bar p$  are available for both the very low energy scattering regime
and the atomic negative energy regime. There are
 two data points for the total annihilation cross section below 3
MeV in the positive energy regime \cite{ZBB99b}, and
three data points for the atomic $2p$ level shift and width plus the
width of the $3d$ level \cite{SBB91} above $-$20 keV in the negative
energy regime. We chose to test the ability of the optical model to
reproduce these data together with the corresponding set of atomic shift
and widths data for $^3$He \cite{SBB91}, although the optical
model approach is not very well suited for studying such light
nuclear targets.

Since potential (F) does not reproduce
very well the $\bar p$-nuclear observables
for $A < 10$, we searched over its parameters to fit
the $\bar p$ -$^{3,4}$He data mentioned above.
As was already encountered in the
fitting procedure described in our earlier work \cite{GFB00} for
potential (a)
of that reference, when trying to fit the
atomic $2p$ and $3d$ levels simultaneously
the calculated width of the $2p$ state in $^4$He came out too
small. This has been noted by other researchers before, as summarized
in Ref. \cite{SBB91}.
We therefore dropped out the $3d$ widths data from the fitting procedure,
partly on the grounds that the $d$-wave contribution to the very low energy
$\bar p$ annihilation
cross sections, which are also being explored here, is almost negligible
compared to the dominant $s$ and $p$ waves (see below).
The resulting density-folded optical potential, referred to as (F'),
has the following parameters: $a_G = 1.8$ fm for the range parameter
of the two-body Gaussian interaction folded in with the He densities,
and a strength parameter $b_0 = - 0.26 + i 2.07$ fm in the notation
of Eq. (\ref{eq:potl}). 
The optical potential (F') is more absorptive
than potential (F) and, again, its real part plays only a minor role.
The $2p$ level shifts and widths, as well as the $\bar p$ total annihilation
cross sections calculated using the optical potential (F') are shown in
Table \ref{tab:He1}, where the measured values are also given,
including a recent report of $\bar p$ annihilation on $^3$He \cite{BBB00c}.
Clearly, potential (F') describes well these data. We note that
omitting the $^3$He data from the fit hardly changes the resultant
(F') which describes very well the $^4$He data. In particular, it was
not possible to decrease, within such a restricted fit, the range
parameter $a_G$ from its relatively large value given above.

Protasov et al. \cite{PBL00} have recently fitted the two low-energy
$\bar p$ - $^4$He total annihilation cross sections listed in the
table, using the SLA expressions for $s$, $p$ and $d$ waves
\cite{CPZ97}. The input $p$- and $d$-wave Coulomb-modified
scattering `lengths' were derived using the Trueman formula \cite{Tru61}
from the $2p$ shift and $2p$ and $3d$ widths. The $s$-wave Coulomb-modified
scattering length $a_{0}^{(sc)}$ was left as a fitting
parameter, since the $\bar p$ atomic $1s$ level
shift and width in He are not known experimentally.
Assuming a value of $1.0 \pm 0.5$ fm for Re $a_{0}^{(sc)}$, these
authors were able to fit the annihilation cross sections
with the following value for the imaginary part of the $s$-wave
Coulomb modified scattering length:
\begin{equation}
\label{eq:scPBL}
{\rm Im}~a^{(sc)}_0 = - 0.36 \pm 0.03 ({\rm stat.})
^{+0.19}_{-0.11}
 ({\rm syst.}) \; {\rm fm} \;\;\; .
\end{equation}
The quality of the fit, as is evident from Fig. 3 in Ref. \cite{PBL00},
is similar to ours (see Table \ref{tab:He1} above), with the
calculated lower (higher) energy cross section somewhat above (below)
the mean value of the measurement. In contrast,
our $\bar p$ - $^4$He potential (F'), which is also fitted to essentially
the same data set, upon using Eq. (\ref{eq:range2})
yields the following values:
\begin{equation}
\label{eq:scG'}
{\rm Re}~a^{(sc)}_0 = 1.851 \; {\rm fm} \; , \;\;\;
{\rm Im}~a^{(sc)}_0 = - 0.630 \; {\rm fm} \;\;\; .
\end{equation}
It is clear from this example
that the `model independent' determination claimed by
Protasov et al. for $a_{0}^{(sc)}$ is violated by our specific example,
suggesting that, contrary to their claim, it is not a model independent
determination. In order to study the origin of the above discrepancy,
we show in Table \ref{tab:He2} the partial wave contributions to
the calculated
cross section at $p_L = 57$ MeV/c.
 The most important contributions are due to  the $s$- and
$p$-waves, for  which the two calculations
give quite different results in spite of agreeing  well with respect
to their sum. In particular, the $p$-wave contributions are quite different
from each other, although sharing practically the $\it same$
value of the Coulomb-modified $p$-wave `scattering length' (more
traditionally called `scattering volume') $a_{1}^{(sc)}$. Whereas
at $-$20 keV, the $l=1$ $\bar p$ - $^4$He dynamics is well
determined by $a_{1}^{(sc)}$ alone, over the energy range of 1 - 3
MeV corresponding to the annihilation measurements, it depends on
more than just  this `scattering length'. The effective
range term, and perhaps higher order terms in the effective range
expansion at low energies, become equally important. Indeed, we have
verified for potential (F') that the variation of the Coulomb-modified
$l=1$ scattering phase shift is not reproduced in this energy range
by specifying the `scattering length' alone. Once the $p$-wave
contributions to the total annihilation cross section differ by as
much as is observed in the table, the essentially fitted $s$-wave
contribution in the calculation of Ref. \cite{PBL00} must also differ,
 and hence our prediction for the Coulomb-modified
$s$-wave scattering length is necessarily different from
that of Ref.\cite{PBL00}.
Such a difference would not occur for the $\bar p p$ system, which
at the appropriate low energies is largely controlled by $s$ waves
\cite{CPZ97}.

Protasov et al. \cite{PBL00} also attached particular significance to
the observation that, for the Coulomb-modified $s$-wave scattering length,
``the absolute value of the scattering length seems to be a decreasing
function of the atomic weight" since ``a naive geometrical picture of
$\bar p$-nucleus annihilation would suggest a value of the
$\bar p$-nucleus scattering length increasing with the nuclear size."
Indeed, their expectation holds for the plain strong-interaction
(not the Coulomb-modified) scattering lengths, as borne out by
the calculations of Ref. \cite{Bat83} which are updated in the next
Section \ref{sec:SLheavy}. However, the Coulomb-modified scattering
lengths do not show such a clear geometrical picture, as is also
discussed in Section \ref{sec:SLheavy}.

\section{Antiproton - nucleus scattering lengths}
\label{sec:SLheavy}

In an earlier paper \cite{Bat83}, hadron nucleus scattering lengths were
derived from exotic atom data for kaons, antiprotons and sigma hyperons. For
the purposes of the present paper, we have updated the previous
calculations for antiprotons, using more recent experimental measurements of
strong interaction shifts and widths.

For exotic atoms with $Z > 1$, strong interaction measurements are only
available for angular momentum states with $l > 0$. As a result it is
necessary to use an optical model approach as an intermediate step in deriving
the $s$-wave Coulomb-modified scattering length $a_{0}^{(sc)}$, or the purely
strong-interaction scattering length $a_0$ from the measured strong interaction
effects for $l > 0$ atomic states. In this method, the optical potential given
by Eq. (\ref{eq:potl}) was used to fit the strong interaction shift and width
values for individual nuclei by adjusting the real and imaginary parts of the
optical strength parameter $b_0$. The $s$-wave scattering length $a_0$
was then calculated
by solving the Klein-Gordon equation for $l=0$ with the optical potential
$V_{\rm opt}$, but without the Coulomb potential, at an energy (1 keV) close to
threshold. Further details of the method are given in the earlier paper
\cite{Bat83}.

The experimental shift and width measurements were taken from the published
literature and cover the available target elements from C to Pr inclusive,
omitting Yb since the nucleus is deformed. The data set used is discussed and
tabulated in \cite{BFG97,BFG95}. For the present work, single particle
distributions were used \cite{BFG97,BFG95} for $\rho(r)$ as these are expected
to be more appropriate for the analysis of antiproton data. In the previous
analysis \cite{Bat83}, macroscopic density distributions were used.

The results of the present analysis are shown in
Fig. \ref{fig:scatl} and are
qualitatively similar to those obtained earlier \cite{Bat83}, where an opposite
sign convention was used for $a_0$ (there denoted $a_s$).
Again the real part of the scattering
length shows an approximate $A^{1/3}$ dependence, and the imaginary part is
constant. This is to be expected on the basis of a simple model based on a
strongly absorbing square well potential
\cite{Bat83}. A least squares fit to these scattering length values gives:

\begin{equation}
\label{eq:sBat}
{\rm Re}~a_0 = (1.54 \pm 0.03)A^{0.311 \pm 0.005} \; {\rm fm} \; , \;\;\;
{\rm Im}~a_0 = -1.00 \pm 0.04 \; {\rm fm} \;\;\; .
\end{equation}
These best fit parameter values are in excellent agreement with those obtained
earlier \cite{Bat83}. We note that the magnitude of Im $a_0$ is considerably
larger than expected for a sharp-edge potential, resulting mainly from the
diffuseness of the potential \cite{Bat83,KPV00}.

For completeness, values for $a_0$ for nuclei with $A < 10$ are also shown in
Fig. \ref{fig:scatl}. The values for $^6$Li and $^7$Li were determined using
the method described above from the strong interaction shift and width
measurements of Poth et al. \cite{PBB87}. The values for $^3$He and $^4$He were
calculated using potential (F') described in Sect. \ref{sec:He}. The value for
hydrogen was obtained with potential (G) described in Sect. \ref{sec:H}. The real
parts of $a_0$ for these light nuclei are seen to deviate from the
predictions of the best fit parameters for $A > 10$ described earlier,
whilst the imaginary parts remain relatively independent of $A$.

To confirm this behaviour for nuclei with $A < 10$, the analysis of the
strong interaction data for C to Pr \cite{BFG97,BFG95}, together with that
for $^6$Li and $^7$Li \cite{PBB87}, was repeated fitting all the data
simultaneously with a single value for the complex parameter $b_0$. A good
fit to the strong interaction data is obtained with a $\chi^2$ of 52 for 46
data points. The $s$-wave scattering lengths obtained in this way are typically a
factor of 5 more precise than those in Fig. \ref{fig:scatl}, which were obtained
using fits to individual nuclei, but they are model dependent since it is
assumed that
the data for all nuclei can be fitted with a single value of $b_0$. However
this seems to be a reasonable assumption in view of the very good fit to the
data. The results confirm the deviation in the value of Re $a_0$ for $^6$Li and
$^7$Li from the simple parameterization for $A > 10$ nuclei shown in
Fig. \ref{fig:scatl}(top). The absolute values of
Im $a_0$ of $1.27 \pm 0.04$ and $1.16 \pm 0.04$ fm for $^6$Li and $^7$Li
respectively, are larger than the mean absolute value of 
Im $a_0 = 1.04 \pm 0.01$ fm
obtained for $A > 10$.

To conclude this section, 
we wish to study the $A$ dependence of the $\bar{p}$-nucleus
Coulomb-modified $s$-wave scattering length $a_{0}^{(sc)}$
for a {\it given} potential, here chosen as potential (F).
The calculated values of $a_{0}^{(sc)}$ for light and medium-weight nuclei
are given in the first two rows of Table \ref{tab:scSL}.
The behaviour of these scattering lengths with $A$ does not follow a
clear geometrical picture. The real part at first increases, it then
decreases very quickly and flips sign, becoming negative already for
$^{12}$C. Another sign flip occurs somewhere between Ne and Ca. This
oscillatory behaviour of Re~$a^{(sc)}_{0}$ is very different from the
smooth rise of Re~$a_{0}$ with $A$ (Eq. (\ref{eq:sBat}) and
Fig. \ref{fig:scatl}). The
magnitude of the imaginary part at first increases with $A$ (rather
than decreasing according to the speculation of Ref. \cite{PBL00}),
reaching a maximum value somewhere near $A \sim 11-12$ approximately
where Re~$a_{0}^{(sc)}$ vanishes, it then decreases to a minimum value
somewhere between Ne and Ca. This behaviour of Im~$a_{0}^{(sc)}$ is
very different from the approximate constancy established above for
Im~$a_{0}$.

The behaviour of $a_{0}^{(sc)}$ as function of $A$ can be
qualitatively understood using the approximate relation
Eq. (\ref{eq:JBl})
between $a_{0}$ and $a_{0}^{(sc)}$ which assumes that $R \ll a_{B}$,
where $R$ is the range of the short-ranged potential giving rise to
$a_{0}$, and $a_{B}$ is the Bohr radius. Neglecting (Im~$a_{0}$)$^{2}$
relative to (Re~$a_{0}$)$^{2}$, approximating $R$ by Re~$a_{0}
\approx 1.5 A^{1/3}$ fm and keeping only the logarithmic term within
the brackets in Eq. (\ref{eq:JBl}),
the condition for Re~$a_{0}^{(sc)}=0$ is
given by $\ln t \approx -t$, where $t = a_{B}^{(p)}/(3A^{4/3})$ with
$a_{B}^{(p)}=57.64$ fm for the Bohr radius of $\bar pp$ and
assuming $Z = A/2$. This condition is satisfied for $A \approx
14$ for which $R = 3.62$~fm according to the above approximations, so
that the condition of applicability $R \ll a_{B} = 4.41$~fm
is still (although barely) satisfied. Relaxing the numerically
unnecessary approximations of neglecting Im~$a_{0}$ and $(1 - 2
\gamma)$, the vanishing of Re~$a^{(sc)}_{0}$ occurs for $A \approx
13.5$, where Im~$a^{(sc)}_{0}$ attains a substantial maximum. Thus,
Eq. (\ref{eq:JBl}) explains well,
within its limited range of validity, the main features of
Re~$a^{(sc)}_{0}$ and Im~$a^{(sc)}_{0}$ from Table \ref{tab:scSL}.
The `non-geometric' dependence of $a^{(sc)}_{0}$ on $A$ is due to
the logarithmic term in Eq. (\ref{eq:JBl})
which arises from the unavoidable
logarithmic behaviour of the Coulomb wavefunctions \cite{Schiff}.

The $s$-wave Coulomb-modified scattering lengths $a_{0}^{(sc)}$
correspond to the short-ranged potential which, strictly speaking, is
superposed on a {\it point} Coulomb potential
$V_{c}^{(point)}= - Z\alpha /r$. This short-ranged potential consists
then, in addition to the strong-interaction optical potential
$V_{\rm opt}$, of the short-ranged Coulomb term
$V_{c}(r) - V_{c}^{(point)}(r)$. In practice we have also included a
vacuum-polarization potential in the latter short-ranged
contribution. The Coulomb-modified phase shifts $\delta _{l}^{(sc)}$
due to the short-ranged part of the interaction, and the related
scattering `lengths' $a_{l}^{(sc)}$ considered in Sect. \ref{sec:opt},
therefore include also effects which are not
entirely due to the strong interactions. To be more precise, the
superscript $(sc)$ should be replaced by $(st + fs; pc)$, that is {\it
strong} + {\it finite size} with respect to {\it point Coulomb}. It
would be more fitting, perhaps, to define the Coulomb-modified phase
shifts $\delta_{l}^{(st; fs+pc)}$, due exclusively to the strong
interactions with respect to the overall Coulomb potential $V_c$.
Similarly, one could also
define the `finite-size' Coulomb-modified phase shifts
$\delta _{l}^{(fs;pc)}$ due to
the short-ranged non strong-interaction potential with respect to
$V_{c}^{(point)}$. It is straightforward to show that

\begin{equation}
\delta _l^{\left( {sc} \right)}\equiv \delta _l^{\left( {st+fs;pc}
\right)}=\delta _l^{\left( {st;fs+pc} \right)}+\delta _l^{\left( {fs;pc}
\right)}\   .
	\label{eq:phases}
\end{equation}
Using Eq. (\ref{eq:range2}) and similarly tailored
low-energy effective-range expansions, Eq. (\ref{eq:phases}) leads
to the following relation between the corresponding scattering lengths:

\begin{equation}
{\tilde{a}}_0^{(sc)} \equiv
a_0^{\left( {st;fs+pc} \right)}={{a_0^{\left( {sc}
\right)}-a_0^{(fs;pc)}} \over {1+\left( {{{2\pi } / {a_B}}}
\right)^2a_0^{\left( {sc} \right)}a_0^{\left( {fs;pc} \right)}}}\   ,
	\label{eq:stSL}
\end{equation}
where $a_0^{(sc)} \equiv {a_0^{\left( {st+fs;pc} \right)}}$ is the same
Coulomb-modified short-ranged scattering length hitherto considered
and tabulated in the first two rows of Table \ref{tab:scSL},
${a_0^{(fs;pc)}}$ is due to the finite size Coulomb (plus vacuum
polarization) effect,
and ${\tilde{a}}_0^{(sc)} \equiv {a_0^{\left( {st;fs+pc} \right)}}$
on the l.h.s. is the intrinsically
strong-interaction Coulomb modified scattering length. We have also
tabulated these newly defined scattering lengths in
Table \ref{tab:scSL}. The real finite size scattering length
$a_0^{(fs;pc)}$ rises modestly with $Z$, becoming
significant for $^{12}$C where it leads to a change
from $-0.6$ fm for Re~$a_{0}^{(sc)}$ to 0.9 fm for
Re~${\tilde{a}}_{0}^{(sc)}$. The induced change for Im~$a_{0}^{(sc)}$
into Im~${\tilde{a}}_{0}^{(sc)}$ already
becomes significant for $^{9}$Be. However, the
general trend of ${{\tilde{a}}_0^{\left( {sc} \right)}}$ with $A$ is
qualitatively similar to that described and discussed earlier for
$a_0^{\left( {sc} \right)}$.

\section{Summary and conclusions}
\label{sec:summ}

We have shown in this work that a unified optical-model
description of low-energy antiproton interactions can be
successfully achieved using a density-folded $\bar p$
optical potential $V_{\rm opt} \sim {\bar v} * {\rho}$.
As shown in Sect. \ref{sec:H},
the potential $\bar v$ gives an excellent fit to the
measured $\bar p p$ annihilation cross sections at low
energies ($p_L < 200$ MeV/c) and to the $1s$ and $2p$
spin-averaged $\bar p$H level shifts and widths, but
is far from being unique; other combinations of range
and of strength parameters have been derived for the
$\bar p p$ system for low as well as for higher energies
\cite{Bru86,BBB00b}, all of which yield a highly
absorptive potential. It would be useful to measure
also $\bar p p$ elastic scattering differential cross
sections at these low energies, including the
Coulomb-nuclear interference angular region,
in order to attempt resolving the ambiguity in the
determination of $\bar v$, particularly its less well
determined real part.

The optical potential $V_{\rm opt}$ reproduces satisfactorily
the $\bar p$ atomic level shifts and widths across the
periodic table for $A > 10$ as well as the few annihilation
cross sections measured on Ne. This simple folding model
does not work well for the very light species of He and Li,
possibly due to the spin and isospin averaging and to other
unjustified approximations inherent in the construction of
optical potentials for these relatively small values
of $A$. However, for a limited range of small-$A$ values,
as shown in Sect. \ref{sec:He} for the He isotopes,
an energy-independent optical potential can be
fitted to both the measured annihilation cross sections
at $E > 0$ and the atomic shift and width data for $E < 0$.
Our detailed analysis of the $\bar p - ^4$He system was
primarily intended to demonstrate the inapplicability of
treating it by the scattering length approximation
\cite{PBL00}. As a byproduct, we have discussed in Sect.
\ref{sec:SLheavy} the systematics
of several sets of $\bar p$ - nucleus $s$-wave scattering
lengths, the purely strong-interaction set for which we
confirmed and updated the discussion of Ref. \cite{Bat83},
and the Coulomb-modified set which defies any
geometrical interpretation.

\begin{figure}
\epsfig{file=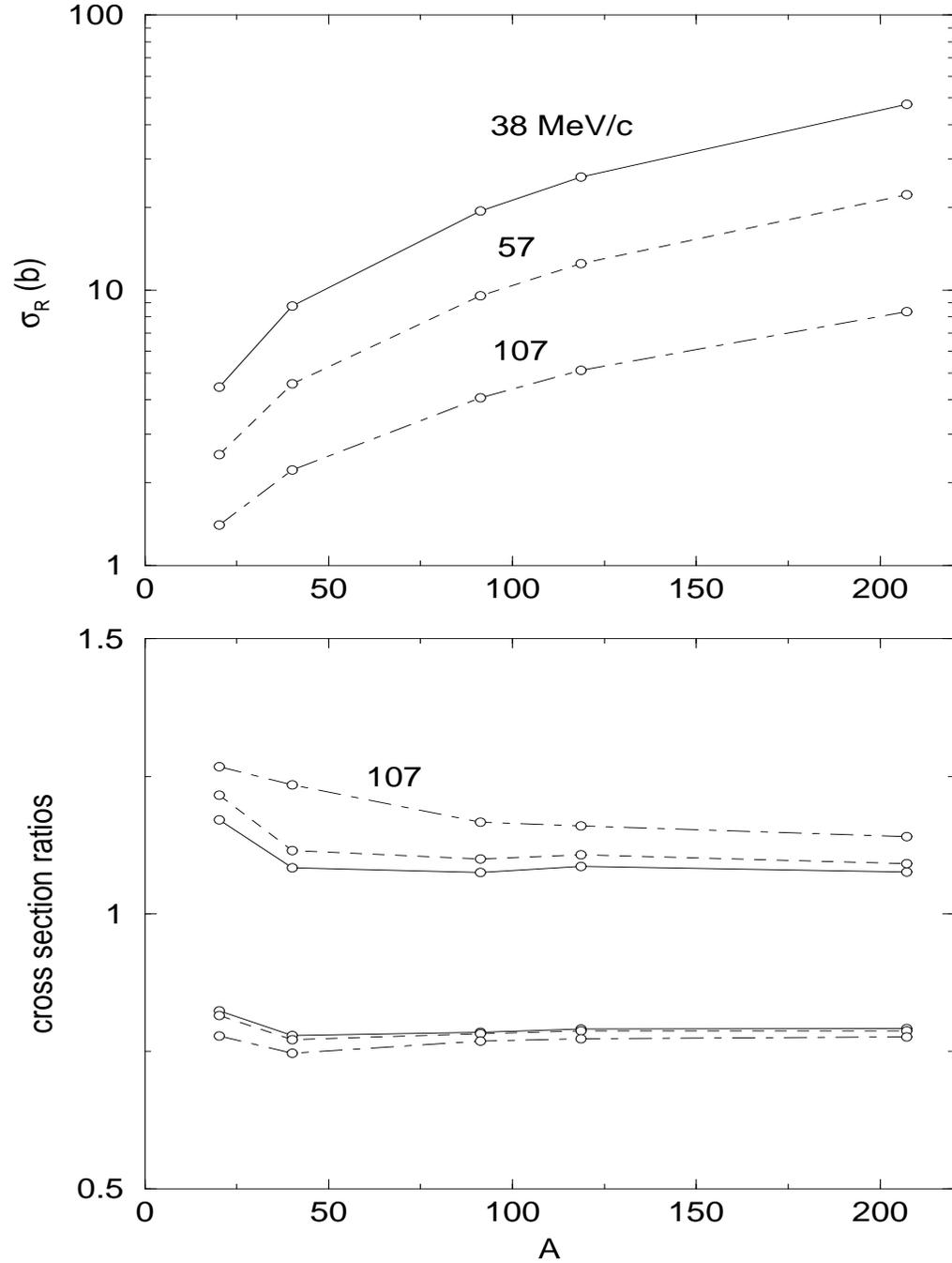,height=180mm,width=135mm,
bbllx=73,bblly=54,bburx=480,bbury=708}
\caption{Top: calculated $\bar p$-nucleus reaction cross sections for
three incoming momenta using the density-folded potential (F).
Bottom: ratios of the above cross sections to the semiclassical
expression Eq. (\ref{eq:sigma}).
Upper band for $R=<r>$, lower band for $R=\frac{4}{3} <r>$.}
\label{fig:heavy}
\end{figure}

\begin{figure}
\epsfig{file=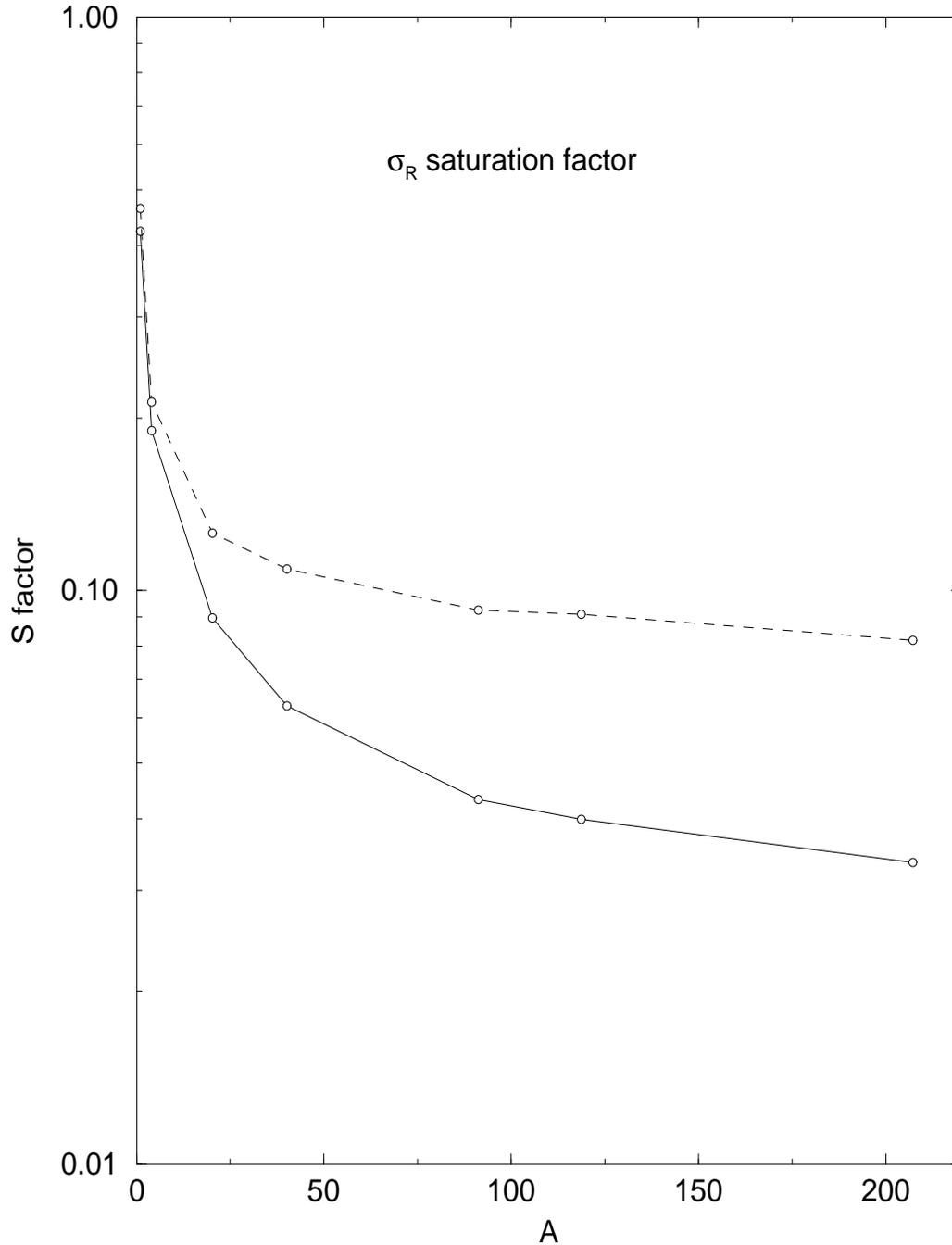,height=180mm,width=135mm,
bbllx=49,bblly=92,bburx=509,bbury=670}
\caption{The saturation factor $S$ of Eq. (\ref{eq:sat})
at $p_L = 57$ MeV/c, calculated using the density-folded
optical potential (F) for $A > 10$, (F') for $A = 4$ and
(G) for $A = 1$, as function of $A$ for antiprotons
(solid line) and for antineutrons (dashed line).}
\label{fig:sat}
\end{figure}

\begin{figure}
\epsfig{file=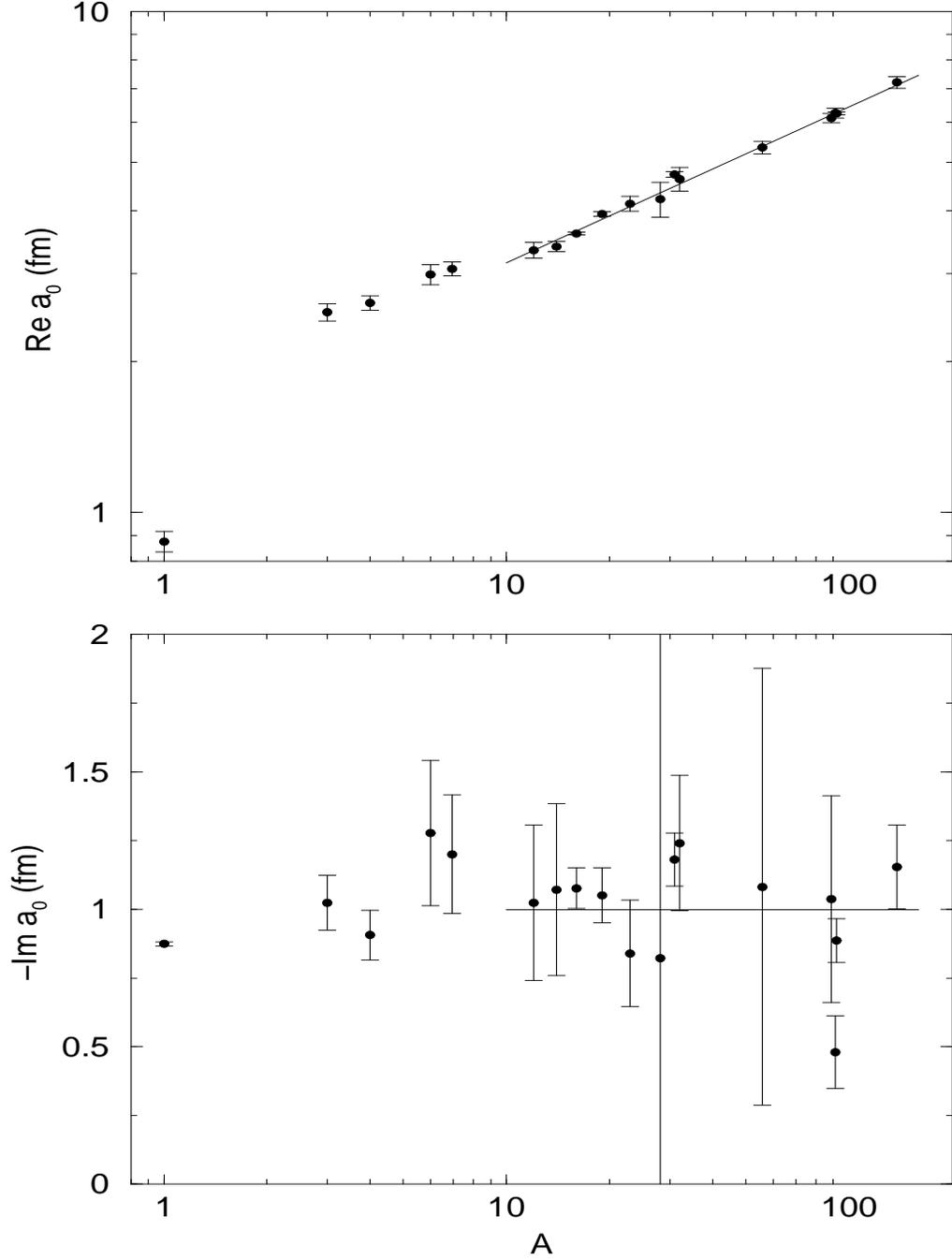,height=180mm,width=135mm,
bbllx=51,bblly=54,bburx=451,bbury=707}
\caption{Real (top) and Imaginary (bottom) parts of the
$s$-wave $\bar p$ scattering length $a_0$ calculated by fitting
to $\bar p$ atomic data.
The straight lines are a best fit to the calculated values 
for $A > 10$, see Eq. (\ref{eq:sBat}).}
\label{fig:scatl}
\end{figure}

Fairly accurate reproduction of the optical potential
calculations of $\sigma _R$ was achieved in
Sect. \ref{sec:heavy} by extending the black-disk
strong-absorption model \cite{Bla54} to account for the
Coulomb focussing effect of negatively charged
projectiles at very low energies ($p_L < 100$ MeV/c).
As the number of additional partial waves contributing to the $\bar p$
reaction cross section
due to the Coulomb focusing effect increases substantially upon
decreasing the incoming energy, the semiclassical description
becomes valid, resulting in the simple closed-form
`geometrical' expression (\ref{eq:sigma}) which embodies the
overall dependence of $\sigma _R$ on $Z$, $A$ and on $E$,
provided the black-disk radius is properly handled
(see Eq. (\ref{eq:rAdep})). Figure \ref{fig:heavy}(bottom)
demonstrates this success for very low $\bar p$ energies
over the periodic table.

A final comment is due to the apparent success of using
a density folded optical potential in terms of a potential
$\bar v$ that describes well the $\bar p p$ system. Deloff
and Law advocated and used a $\bar v * \rho$ folding model
for kaonic atoms \cite{DLa74a,Del80} and for $\bar p$ atoms
\cite{DLa74b}. This model was subsequently
tested by Batty for kaonic atoms \cite{Bat81a} and for
antiprotonic and sigma atoms \cite{Bat81b}, and moderately
good agreement with the data was found.
However, the underlying potentials $\bar v$
were not fitted to a comprehensive set of $Z = 1$ data.
The present work provides a step forward for antiprotons,
in that a complete body of $\bar p p$ data was fitted
to determine $\bar v$. This has to be supplemented in due
course by $\bar p n$ data, particularly in order to understand
the renormalization factor of 2/3 encountered in obtaining
the density-folded optical potential (F) from the $\bar v$
potential (G).
A multiple-scattering justification of the
$\bar v * \rho$ folding model was given by Green and Wycech
\cite{GWy82} for antiprotons as due to the strong
absorptivity, and was confirmed in detailed calculations
by other authors \cite{SNa84,KGE84}.

Clearly, more data on $\bar p$ nuclear interactions
at low energies are needed to develop quantitatively
further the unified optical-model approach put forward
in the present work.
Such data should consist of elastic and charge exchange
scattering, as well as annihilation cross sections on
nuclei. In particular, data on the very light nuclei,
for $A < 10$, would provide valuable information on how
to bridge the gap, prevailing at present within the
density-folded optical model approach, between hydrogen
and the $A > 10$ nuclei  concerning the applicability of
the same (smoothly dependent on $A$) $\bar p$ optical
potential.

\vspace{10mm}

CJB wishes to thank the Hebrew University for support for a visit
during which this work was started.
\newline
This research was partially supported by the Israel Science Foundation.

\begin{table}
\caption{Measured and calculated annihilation cross
sections for $\bar p$ on Ne. Calculations were made with potential (F).}
\label{tab:Ne}
\begin{tabular}{ccc}
$p_L$ (MeV/c)&$\sigma ^{(\rm exp)}_{ann}$ (mb)
&$\sigma ^{(\rm calc)}_{ann}$ (mb) \\ \hline
57&2210$\pm1105 ^a$&2524\\
192.8&956$\pm47 ^b$&1016\\
\end{tabular}
\end{table}
\noindent
$^a$ Ref.\cite{BBB00a} \newline
$^b$ Ref.\cite{BBB86} \newline

\begin{table}
	\caption{Measured and calculated observables for the
  $\bar p  - ^{3,4}$He system. Shifts ($\varepsilon$) and
  widths ($\Gamma$) are in eV,
  cross sections are in mb. Calculations were made with potential (F').}
\label{tab:He1}
%\protect\label{tab:He1}
\begin{tabular}{lcccccc}
\hline
& & $\varepsilon _{2p}$ & $\Gamma_{2p}$ &
 $\sigma_{ann}$ (47.0 MeV/c)  &
$\sigma_{ann}$ (55.0 MeV/c)& $\sigma_{ann}$ (70.4 MeV/c)  \\
\hline
$^{3}$He & calc.& $-12$ & 33 & --- & 1038 &--- \\

$^{3}$He & exp.&$-17\pm 4^a$&$25\pm9 ^a$& --- & 1850$\pm 700^c$& ---  \\
	
$^4$He&calc. & $-19$ &42& 1116 & --- &810  \\

$^4$He&exp. & $-18 \pm 2^{a}$ & $45 \pm 5^{a}$& $979 \pm
145^{b}$ & ---& $827 \pm 38^{b}$  \\
\hline
\end{tabular}
\end{table}
\noindent
$^{a}$ Ref. \cite{SBB91} \\
$^{b}$ Ref. \cite{ZBB99b} \\
$^{c}$ Ref. \cite{BBB00c}
%\newpage

\begin{table}
\caption{Partial wave contributions (in mb) to
the calculated $\bar p$ - $^{4}$He total
annihilation cross section at $p_L=57$ MeV/c.}
\label{tab:He2}
%\protect\label{tab:He2}
	\begin{tabular}{lccccc}
		\hline
  & $l = 0$ & $l = 1$ & $l = 2$&$l = 3$ & sum  \\
		\hline
Protasov et al.$^{a}$ & 280.3 & 652.5 & 16.2 &  & 949  \\
potential (F') & 395.8 & 500.3 & 49.8 & 1.5 & 949  \\
experiment$^{b}$ &  &  &  &  & $915 \pm 32$  \\
\hline
\end{tabular}
\end{table}
\noindent
$^{a}$ Ref. \cite{PBL00} \\
$^{b}$ Ref. \cite{BBB00a}	
%\end{table}

\begin{table}
\caption{Coulomb-modified $s$-wave $\bar p$
scatttering lengths (in fm)
	calculated for potential (F).}
\label{tab:scSL}
%\protect\label{tab:scSL}
\begin{tabular}{lcccccccc}
\hline
target & $^{4}$He & $^{6}$Li & $^{9}$Be & $^{10}$B & $^{12}$C &
$^{16}$O & Ne & Ca  \\
\hline
Re $a_{0}^{(sc)}$ & 1.562 & 1.967 & 2.169 & 1.401 & $-0.626$ &
$-0.645$ & $-0.216$ & 0.134  \\
Im $a_{0}^{(sc)}$ & $-0.590$ & $-1.012$ & $-1.792$ &$-2.853$ &$-2.631$ &
$-0.551$ & $-0.241$ & $-0.391$    \\
$a_{0}^{(fs;pc)}$ & 0.067 & 0.148 & 0.170 & 0.183 & 0.176 & 0.252 &
0.307 & 0.835 \\
Re $\tilde a_{0}^{(sc)}$ & 1.479 & 1.703 & 1.809 & 1.633 & 0.948 &
$-0.654$ & $-0.485$ & 0.004 \\
Im $\tilde a_{0}^{(sc)}$ & $-0.576$ & $-0.853$ & $-1.183$ &
$-1.612$ & $-2.373$ & $-1.414$ & $-0.556$ & $-0.121$ \\
\hline
\end{tabular}
\end{table}

\end{document}